\documentclass[mathpazo]{cicp}
\usepackage[dvips]{graphicx}
\newcommand{\la}{\left\langle}
\newcommand{\ra}{\right\rangle}
\newcommand{\EPL}{Europhys.~Lett.~}
\newcommand{\PRL}{Phys.~Rev.~Lett.~}
\newcommand{\PR}{Phys.~Rev.~}
\newcommand{\JCP}{J.~Chem.~Phys.~}

\newcommand{\JPCM}{J.~Phys.: Condens.~Matter~}
\newcommand{\MP}{Mol.~Phys.~}
\newcommand{\JCIS}{J.~Coll.~Int.~Sci.~}
\newcommand{\EPJ}{Eur.~Phys.~J.~}

\begin{document}

\title{Charge Renormalization, Thermodynamics, and \\
Structure of Deionized Colloidal Suspensions}

\author{Ben Lu and Alan R. Denton\corrauth}
\address{Dept.~of Physics, North Dakota State University, Fargo, ND, U.S.A. 58108-6050}
\email{{\tt alan.denton@ndsu.edu}}

\date{\today}

\begin{abstract}
In charge-stabilized colloidal suspensions, highly charged macroions, dressed by
strongly correlated counterions, carry an effective charge that can be substantially
reduced (renormalized) from the bare charge.  Interactions between dressed macroions
are screened by weakly correlated counterions and salt ions.  Thermodynamic
and structural properties of colloidal suspensions depend sensitively on the
magnitudes of both the effective charge and the effective screening constant.
Combining a charge renormalization theory of effective electrostatic interactions
with Monte Carlo simulations of a one-component model, we compute osmotic pressures
and pair distribution functions of deionized colloidal suspensions.  This
computationally practical approach yields close agreement with corresponding results
from large-scale simulations of the primitive model up to modest electrostatic
coupling strengths.
\end{abstract}


\maketitle

\section{Introduction}\label{intro}

Charge-stabilized colloidal suspensions~\cite{Evans,Schmitz,Pusey} exhibit rich
thermodynamic phase behavior and tunable materials properties (e.g., thermal,
mechanical, optical, rheological) that are the basis of many industrial and
technological applications.
Thermally excited (Brownian) motion of nm-$\mu$m-sized particles dispersed in a
fluid medium drives the self-assembly of ordered phases.  Important examples are
nanoscale structures~\cite{zvelindovsky07} and colloidal crystals, whose diverse
crystalline symmetries and variable lattice constants can conveniently template
photonic band-gap materials~\cite{photonics1,photonics2,photonics3}.

Interparticle interactions and correlations determine the distribution of microions
(counterions and salt ions) around the colloidal macroions and thereby govern
microion-mediated electrostatic interactions between macroions.  As explained by
the landmark theory of Derjaguin, Landau, Verwey, and Overbeek (DLVO)~\cite{DL,VO},
repulsive interactions between like-charged macroions can stabilize a
suspension against aggregation induced by van der Waals attractive
interactions~\cite{Israelachvili,interactions}.  Equilibrium and dynamical properties
of charged colloids depend sensitively on electrostatic interactions, which can be
widely tuned by adjusting system parameters: size, charge, and concentration of
ion species; pH and dielectric constant of solvent.

A powerful approach to modeling charged colloids is molecular simulation of either
the primitive model, which includes all ions explicitly, or all-atom models,
which include even individual solvent (e.g., water) molecules.
At such microscopic resolution, simulations can potentially yield valuable insights
into the thermodynamic, structural, and dynamical properties of real suspensions.
Currently available processors and algorithms are limited, however, to relatively
small systems and low ion size and charge ratios.  Furthermore, simulations of
microscopically detailed models do not necessarily elucidate the physical mechanisms
underlying complex cooperative behavior.

An alternative to brute-force modeling is simulation of a one-component model,
derived by pre-averaging over the microion coordinates to obtain {\it effective}
interactions between macroions, screened by implicitly modeled microions.
This coarse-grained strategy finesses the computational challenges that plague
more explicit models, but relies on practical and accurate approximations for
the effective interactions.  A previous simulation study~\cite{lu-denton07}
validated the one-component model --- implemented with linear-response and mean-field
approximations --- by direct comparison with pressure data from primitive model
simulations at modest electrostatic couplings.  The eventual breakdown of the model
at higher couplings was attributed to failure of linear-screening approximations to
account for nonlinear counterion association near macroions.

A recently proposed charge renormalization theory of effective interactions in
charged colloids~\cite{denton08} addresses limitations of linear-screening
approximations by incorporating an effective (renormalized) macroion charge into
the one-component model.  Calculations based on a variational approximation for
the free energy indicated the potential of the theory to accurately predict
thermodynamic properties of suspensions well into the nonlinear-screening parameter
regime.  The present paper describes complementary Monte Carlo simulations designed
to test predictions for both thermodynamics (osmotic pressure) and structure
(pair distribution function) of deionized suspensions of highly charged colloids.

The remainder of the paper is organized as follows.  Section~\ref{models} first
reviews the primitive and effective one-component models of charged colloids.
Section~\ref{theory} outlines the charge renormalization theory, which predicts
renormalized system parameters: effective macroion charge, volume fraction,
and screening constant.  Section~\ref{simulations} describes our Monte Carlo
simulations, which take as input the renormalized effective interactions.
Section~\ref{results} compares our results for the pressure and pair distribution
function of deionized suspensions with corresponding data from simulations of
the primitive model and with predictions of a variational free energy theory.
Excellent agreement is obtained, with minimal computational effort, over ranges
of macroion charge, volume fraction, and electrostatic coupling strength.
Finally, Sec.~\ref{conclusions} closes with a summary and conclusions.

\section{Models}\label{models}

The primitive model idealizes the colloidal macroions as charged hard spheres
of monodisperse radius $a$ and bare valence $Z_0$ (charge $-Z_0e$), the microions
as point charges, and the solvent as a continuum of uniform dielectric constant
$\epsilon$.  We limit present consideration to monovalent microions, for which
microion-microion correlations are relatively weak, and further assume index-matching
of macroions and solvent, justifying neglect of image charges and polarization
effects.  At absolute temperature $T$, a characteristic length scale is the
Bjerrum length $\lambda_B=e^2/(\epsilon k_BT)$, defined as the distance between
a microion pair at which the electrostatic interaction energy rivals
the average thermal energy.  In a closed volume $V$, the suspension comprises $N_m$
macroions and $N_{\pm}$ positive/negative microions, of which $N_c$ are counterions
and $N_s$ are salt ion pairs.  Global electroneutrality relates the ion numbers via
$N_c=N_+-N_-=Z_0N_m$.  Alternatively, the suspension may be in Donnan equilibrium
(e.g., across a semi-permeable membrane) with an electrolyte reservoir that fixes
the salt chemical potential.

While the primitive model has the virtue of explicitly representing all ions,
simulations face severe computational challenges posed by long-range Coulomb
interparticle interactions and large size and charge asymmetries between ion species.
Sophisticated numerical algorithms, such as cluster moves~\cite{lobaskin-linse99,luijten}
and Ewald or Lekner summation~\cite{frenkel,allen}, can significantly broaden the
range of accessible length and time scales.  Nevertheless, currently feasible studies
are still limited to relatively small systems and modest asymmetries.

An alternative approach to modeling charged colloids is derived from the
Hamiltonian $H$ of the primitive model by formally tracing over the degrees
of freedom of the microions ($\mu$) in the partition function of the
multi-component mixture:
\begin{equation}
{\cal Z}=\langle\langle\exp(-H/k_BT)\rangle_{\mu}\rangle_m
=\la\exp(-H_{\rm eff}/k_BT)\ra_m~,
\label{part2}
\end{equation}
leaving an explicit trace over only the macroion ($m$) degrees of freedom.
The resulting one-component model (OCM) is governed by an effective Hamiltonian
\begin{equation}
H_{\rm eff}=E+\frac{1}{2}\sum_{i\neq j=1}^{N_m}v_{\rm eff}(|{\bf r}_i-{\bf r}_j|)
+\cdots~,
\label{Heff}
\end{equation}
a function of the macroion coordinates ${\bf r}_i$, which comprises a one-body
volume energy $E$, an effective macroion-macroion pair potential $v_{\rm eff}(r)$,
and higher-order terms involving sums over effective many-body potentials.
The volume energy, although independent of macroion coordinates, depends on the
average densities of macroions and salt ions.  The effective pair potential arises
from screening of the bare Coulomb potential by the implicitly modeled microions.

The computational benefit of reducing the number of components and the interaction
range comes at a cost of increased complexity in the effective interactions.
Linear-screening and mean-field approximations, which are justifiable in the case
of weakly correlated monovalent microions, yield analytical forms for the
volume energy and effective pair potential~\cite{silbert91,denton99,denton00}.
Moreover, many-body effective interactions are usually weak~\cite{denton04,denton06}.
The range of validity of the conventional OCM is nevertheless limited by the
inevitable onset of nonlinear screening with strengthening macroion-counterion
correlations.  Previous studies~\cite{lu-denton07,denton06} establish a parameter
threshold $Z_0\lambda_B/a\simeq 7$ --- characteristic of highly charged latex
particles and ionic surfactant micelles --- above which nonlinear effects
tend to become significant.

\begin{figure}
\begin{center}
\includegraphics[width=0.6\columnwidth]{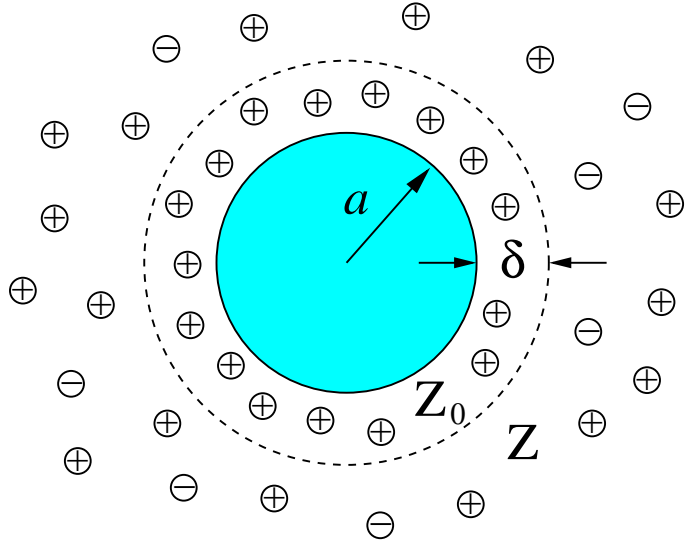}
\end{center}
\caption{\label{fig-model}
Model of charged colloidal suspension: spherical macroions of radius $a$
and point microions dispersed in a dielectric continuum.  Strongly associated
counterions in a spherical shell of thickness $\delta$ renormalize the
bare macroion valence $Z_0$ to an effective (lower) valence $Z$.
}
\end{figure}

The OCM can be extended to more strongly correlated systems by incorporating
physical concepts from charge renormalization theory~\cite{alexander84}.
Counterions that venture sufficiently close to a highly charged macroion
may become thermally ``bound" --- closely associated with the macroion, if not
condensed onto its surface~\cite{manning69,Oosawa}.  As Fig.~\ref{fig-model} depicts,
counterions localized within a spherical shell surrounding the macroion may be
considered to renormalize the bare macroion valence.  The composite ``dressed"
macroion, comprising a bare macroion and its shell of bound counterions,
carries a reduced effective valence $Z\le Z_0$.

Counterions bind to a macroion at a distance at which the electrostatic energy of
attraction is comparable to the average thermal energy per counterion.  Denoting
by $\phi(r)$ the electrostatic potential at distance $r$ from a macroion center,
the width of the association shell $\delta$ can be defined via
\begin{equation}
e|\phi(a+\delta)|=Ck_BT~,
\label{delta1}
\end{equation}
where $C$ is a dimensionless parameter of order unity.  In general, the reference
potential should be chosen as the Donnan (average) potential $\phi_D$ of the
suspension, $C$ then being a function of salt concentration.  Counterions within
an association shell ($a<r<a+\delta$) are presumed bound by the potential well
of the respective macroion, while more distant counterions roam free and
screen the dressed macroions.  In other studies, a similar thermal criterion has
been applied to the electrostatic potential~\cite{schmitz99,schmitz00,schmitz02}
and the effective pair potential~\cite{safran00}.

\section{Theory}\label{theory}

The effective interaction theory proposed in ref.~\cite{denton08} combines
a charge renormalization theory for the effective macroion charge with a
mean-field, linear-response theory of the OCM~\cite{silbert91,denton99,denton00}.
Here we briefly outline the theory, whose predictions for effective interactions
are subsequently used as input to our simulations (Sec.~\ref{simulations}).

By invoking a mean-field random-phase approximation for the response functions
(partial static structure factors) of a microion plasma, linear-response theory
yields analytical expressions for effective electrostatic interactions.
This approach proves formally equivalent to density-functional~\cite{vRH97,graf98,
vRDH99,vRE99,zoetekouw_pre06}, extended Debye-H\"uckel~\cite{warren}, and
distribution-function~\cite{chan85,chan01} formulations of effective interaction
theory~\cite{zvelindovsky07}.  Linear-response theory is also equivalent to
linearized Poisson-Boltzmann (PB) theory, based on a consistent linearization
of the PB equation and expansion of the ideal-gas free energy functional
to quadratic order in the microion density profiles~\cite{denton07}.

Linearized theories predict the reduced electrostatic potential generated by
a single bare macroion as
\begin{equation}
\psi(r)\equiv\beta\phi(r)=-Z_0\lambda_B~\frac{e^{\kappa a}}{1+\kappa a}
~\frac{e^{-\kappa r}}{r}~, \quad r\ge a~,
\label{psi0}
\end{equation}
where $\beta=1/k_BT$, $\kappa=\sqrt{4\pi\lambda_B(n_++n_-)}$ denotes the bare Debye
screening constant, and $n_{\pm}=N_{\pm}/[V(1-\eta)]$ are the mean number densities
of microions in the free volume, i.e., the total volume $V$ reduced by the fraction
$\eta$ occupied by the macroion hard cores.  The constraint of global
electroneutrality ensures that $\kappa$ depends implicitly on the average densities
of both macroions and salt ions.

Adapting Eq.~(\ref{psi0}) to the charge renormalization model (Fig.~\ref{fig-model}),
the potential around a dressed macroion, of effective valence $Z$ and effective
radius $a+\delta$, becomes
\begin{equation}
\tilde\psi(r)=-Z\lambda_B~\frac{e^{\tilde\kappa(a+\delta)}}{1+\tilde\kappa(a+\delta)}
~\frac{e^{-\tilde\kappa r}}{r}, \quad r\ge a+\delta~,
\label{psir}
\end{equation}
where $\tilde\kappa=\sqrt{4\pi\lambda_B(\tilde n_++\tilde n_-)}$ now represents the
effective (renormalized) screening constant,
$\tilde n_{\pm}=\tilde N_{\pm}/[V(1-\tilde\eta)]$ and $\tilde N_{\pm}$ are the
mean number densities and numbers of free microions,
and $\tilde\eta=\eta(1+\delta/a)^3$ is the effective volume fraction
of the dressed macroions.  Combining Eqs.~(\ref{delta1}) and (\ref{psir}) yields the
transcendental equation
\begin{equation}
\frac{Z\lambda_B}{[1+\tilde\kappa(a+\delta)](a+\delta)}=C~,
\label{delta2}
\end{equation}
which may be solved for the association shell thickness $\delta$ for given values
of $Z$, $\eta$, and $C$, noting that $\tilde\kappa$ depends self-consistently on
$\delta$.

The distinction between free and bound microions implies a corresponding separation
of the total free energy
\begin{equation}
F=F_{\rm free}+F_{\rm bound}+F_m~,
\label{Ftot}
\end{equation}
into three terms representing, respectively, contributions from free and bound
microions and from macroion effective interactions.  The free microions are,
by construction, sufficiently weakly correlated with the macroions to be well
described by linear-response theory.  Their free energy (per macroion) is
therefore accurately approximated by the linear-screening prediction for the
volume energy (with renormalized parameters):
\begin{equation}
\beta f_{\rm free}=\sum_{i=\pm}\tilde x_i[\ln(\tilde n_i\Lambda^3)-1]
-\frac{Z^2}{2}~\frac{\tilde\kappa\lambda_B}{1+\tilde\kappa(a+\delta)}
-\frac{Z^2}{2}~\frac{n_m}{\tilde n_++\tilde n_-}~,
\label{ffree}
\end{equation}
where $\tilde x_{\pm}=\tilde N_{\pm}/N_m$ are the free microion concentrations,
$\Lambda$ is the microion thermal de Broglie wavelength, and $n_m=N_m/V$ is the
mean macroion number density.  The three terms on the right side have physical
interpretations as the ideal-gas (entropic) free energy of the free microions,
the self-energy of the dressed macroions, and the Donnan (average) potential
energy of the microions.
Although the bound counterions, being relatively strongly correlated with the
macroions, do not yield to a linear-response treatment, their free energy
(per macroion) can be reasonably approximated by
\begin{equation}
\beta f_{\rm bound}=(Z_0-Z)\left[\ln\left(\frac{Z_0-Z}{v_s}\Lambda^3
\right)-1\right]+\frac{Z^2\lambda_B}{2a}~,
\label{fb}
\end{equation}
where the first term on the right side is an ideal-gas contribution, $(Z_0-Z)/v_s$
being the mean counterion density in the association shell of volume
$v_s=(4\pi/3)[(a+\delta)^3-a^3]$, and the second term is the electrostatic energy,
assuming tight localization of the counterions near $r=a$.

The effective valence $Z$ is now prescribed, for a given bare valence $Z_0$,
by the variational ansatz~\cite{levin98,tamashiro98,levin01}
\begin{equation}
\left(\frac{\partial}{\partial Z}(f_{\rm free}+f_{\rm bound})
\right)_{T,n_{\pm}}=0~,
\label{Z}
\end{equation}
which ensures equality of the chemical potentials of free and bound counterions,
under the constraint that $Z$ and $\delta$ are connected by Eq.~(\ref{delta2}).
The effective valence and shell thickness then determine the effective screening
constant $\tilde\kappa$.

The macroion free energy $F_m$ in Eq.~(\ref{Ftot}) depends on correlations and
effective interactions between dressed macroions.  Incorporating renormalized
system parameters into linear-response theory~\cite{denton99,denton00},
the effective pair potential is given by
\begin{equation}
\beta v_{\rm eff}(r)=Z^2\lambda_B\left(\frac{e^{\tilde\kappa a}}
{1+\tilde\kappa a}\right)^2\frac{e^{-\tilde\kappa r}}{r}~,\quad r\ge 2(a+\delta)~,
\label{veffr}
\end{equation}
whose screened-Coulomb form is identical to the long-range limit of the DLVO
pair potential~\cite{VO}.
The renormalized effective interactions and system parameters can be input into
computer simulations or liquid-state theories~\cite{HM} of the OCM to compute
thermodynamic and structural properties of bulk suspensions of charged colloids.
In the next section, we compare our simulation results with predictions of a
variational approximation~\cite{denton06,vRH97} for the macroion free energy
(per macroion) based on first-order thermodynamic perturbation theory with a
hard-sphere reference system:
\begin{equation}
f_m(n_m,\tilde n_{\pm})=\min_{(d)}\left\{f_{\rm HS}(n_m,\tilde n_{\pm};d)
+2\pi n_m\int_d^{\infty}{\rm d}r\,r^2g_{\rm HS}(r,n_m;d)
\tilde v_{\rm eff}(r,n_m,\tilde n_{\pm})\right\}~.
\label{fm}
\end{equation}
Here $f_{\rm HS}$ and $g_{\rm HS}$ are, respectively, the excess free energy
density and pair distribution function of the HS fluid, computed from the
near-exact Carnahan-Starling and Verlet-Weis expressions~\cite{HM}.
Minimization with respect to the effective hard-sphere diameter $d$ yields
a least upper bound to the free energy.  In practice, the renormalized system
parameters ($Z$, $\delta$, $\tilde\kappa$) must be held fixed in this
minimization and in all partial thermodynamic derivatives.
The corresponding prediction for the thermodynamic pressure is
\begin{equation}
\beta p=n_m+\tilde n_++\tilde n_-
-\frac{Z(\tilde n_+-\tilde n_-)\tilde\kappa\lambda_B}{4[1+\tilde\kappa(a+\delta)]^2}
+n_m^2\beta\left(\frac{\partial f_m}{\partial n_m}\right)_{T,N_s/N_m}~,
\label{ptot}
\end{equation}
where the first three terms on the right side are ideal-gas contributions from
macroions and free microions, the fourth term arises from density dependence of
the self-energy of the dressed macroions, and the final term is generated by
effective interactions between pairs of dressed macroions.
While this simple variational theory yields the pressure and other
thermodynamic properties, computer simulations or integral-equation theory
are required to determine structural properties.

\section{Monte Carlo Simulations}\label{simulations}

Working within the canonical (constant-$NVT$) ensemble, we consider a one-component
fluid of macroions in a cubic box, subject to periodic boundary conditions, at fixed
temperature, volume, macroion number, and mean salt concentration.
The macroions interact with one another via effective electrostatic interactions
that depend on the mean densities of both macroions and (implicitly modeled) salt.
According to the standard Metropolis algorithm~\cite{frenkel,allen}, trial particle
displacements are accepted with probability
\begin{equation}
P=\min\left\{\exp(-\beta\Delta U),1\right\}~,
\label{acc}
\end{equation}
where $\Delta U=U(n)-U(o)$ is the change in pair potential energy between
the new (n) and old (o) states and
\begin{equation}
U=\sum_{i<j=1}^{N_m}v_{\rm eff}(|{\bf r}_i-{\bf r}_j|)
\label{U}
\end{equation}
is a sum of hard-sphere-repulsive-Yukawa (screened-Coulomb) pair potentials.
Note that the acceptance probability for trial displacements does not involve
the volume energy, since the mean density is fixed.  Consequently, the volume energy
does not affect the macroion structure, although it does contribute to the pressure
(see below).  To achieve high numerical precision, pair interactions were cut off
at a distance $r_c\simeq 20/\tilde\kappa$, i.e., 20 effective screening lengths.
The cut-off radius determined the minimum box side length, $L=2r_c$, necessary to avoid
interactions of a particle with its own periodic images.  For a given volume fraction,
the requisite number of macroions was therefore prescribed as
$N_m\simeq (3\eta/4\pi)(40/\tilde\kappa a)^3$.

We performed a series of simulations, using renormalized effective interaction
parameters (effective macroion valence, volume fraction, and screening constant),
starting from face-centered cubic crystal configurations.  Trial moves were executed
by randomly displacing macroions with a step size adjusted to yield an acceptance
rate of about 50\%.  Following an equilibration phase of $10^4$ cycles, statistics
were accumulated for pressures and pair distribution functions over the next $10^4$
cycles ($N_m\times 10^4$ particle displacements).  Test simulations for larger
systems confirmed finite-size effects to be negligible.  Relatively modest computing
resources were required, with typical runs on a single Intel Xeon-HT processor
lasting 30, 120, and 200 hours for $N_m=2400$, 4000, and 5800 particles, respectively.

In the constant-$NVT$ ensemble, the total pressure may be expressed as
\begin{equation}
\beta p=n_m+\tilde n_++\tilde n_-
-\frac{Z(\tilde n_+-\tilde n_-)\tilde\kappa\lambda_B}{4[1+\tilde\kappa(a+\delta)]^2}
+\beta p_{\rm pair}~,
\label{pvirial}
\end{equation}
where $p_{\rm pair}$ is generated by effective interactions between pairs of
macroions and corresponds to the last term on the right side of Eq.~(\ref{ptot}).
The pair pressure is computed from the virial expression for a
density-dependent pair potential~\cite{louis02}:
\begin{equation}
p_{\rm pair}=\la\frac{{\cal V}_{\rm int}}{3V}\ra
-\la\left(\frac{\partial U}{\partial V}\right)_{\tilde N_s/N_m}\ra
+p_{\rm tail}~,
\label{ppair}
\end{equation}
where ${\cal V}_{\rm int}$ is the internal virial, the volume derivative term
accounts for the density dependence of the effective pair potential, angular
brackets denote an ensemble average over configurations in the canonical ensemble,
and $p_{\rm tail}$ corrects for cutting off the long-range tail of the pair potential.
The internal virial is given by
\begin{equation}
{\cal V}_{\rm int}=\sum_{i=1}^{N_m}{\bf r}_i\cdot{\bf f}_i
=\sum_{i<j=1}^{N_m}(1+\tilde\kappa r_{ij})v_{\rm eff}(r_{ij})~,
\label{virial}
\end{equation}
where ${\bf f}_i=-\sum_{j\neq i}v'_{\rm eff}(r_{ij})$ is the effective
force exerted on macroion $i$ by all neighboring macroions within a sphere
of radius $r_c$.  The second term on the right side of Eq.~(\ref{ppair}) is
computed as the ensemble average of
\begin{equation}
\left(\frac{\partial U}{\partial V}\right)_{\tilde N_s/N_m}
=-\frac{n_m}{V}\sum_{i<j=1}^{N_m}\left(
\frac{\partial v_{\rm eff}(r_{ij})}{\partial n_m}\right)_{\tilde N_s/N_m}
=\frac{1}{V(1-\tilde\eta)}\sum_{i<j=1}^{N_m}
\left(\frac{\tilde\kappa r_{ij}}{2}-\frac{\tilde\kappa^2 a^2}{1+\tilde\kappa a}\right)
v_{\rm eff}(r_{ij})~.
\label{dUdV}
\end{equation}
Finally, neglecting pair correlations for $r>r_c$, the tail pressure is
approximated by
\begin{equation}
p_{\rm tail}=-\frac{2\pi}{3}n_m^2\int_{r_c}^\infty{\rm d}r\,r^3
v'_{\rm eff}(r)=\frac{2\pi}{3}n_m^2
\left(\frac{\tilde\kappa^2 r_c^2+3\tilde\kappa r_c+3}{\tilde\kappa^2}\right)
r_c v_{\rm eff}(r_c)~.
\label{ptail}
\end{equation}

The macroion structure of the suspension is characterized by the macroion-macroion
pair distribution function $g(r)$, defined such that $4\pi r^2g(r){\rm d}r$ equals
the average number of macroions in a spherical shell of radius $r$ and thickness
${\rm d}r$ centered on a macroion.  With each particle regarded in turn as the
central particle in a given configuration, neighboring particles were assigned to
concentric spherical shells (bins), of thickness $\Delta r=0.1 a$, according to
their radial distance $r$ from the central particle.  Following equilibration,
$g(r)$ was computed in the range $2(a+\delta)<r<L/2$ by accumulating statistics of
particle numbers in radial bins and averaging over all configurations.  The raw
distributions were finally smoothed by averaging each bin together with its neighbors
in a moving average algorithm.

\section{Results and Discussion}\label{results}

The renormalized effective interactions (Sec.~\ref{theory}) previously provided
the basis for variational theory calculations of the pressure of deionized
suspensions~\cite{denton08}.  The same effective interactions are here input into
Monte Carlo simulations (Sec.~\ref{simulations}) of the one-component model to
compute both the pressure and macroion-macroion pair distribution function.
Comparing our results with corresponding data from primitive model simulations
allows testing the accuracy of the effective interactions and the variational
free energy approximation.

All results presented below are for the case of monovalent counterions, zero salt
concentration, and an aqueous solvent at room temperature ($\lambda_B=0.714$ nm).
Throughout, the dimensionless thermal parameter in Eq.~(\ref{delta2}) is fixed at
$C=3$, a value shown in ref.~\cite{denton08} to give satisfactory agreement with
pressure data from primitive model simulations.  For salt-free suspensions, this
choice of $C$ corresponds to $e|\phi(a+\delta)-\phi_D|=2k_BT$.  In passing, we note
that the parameter $C$ is analogous to the adjustable cell radius $b$ in the PB 
cell theory of Zoetekouw and van Roij~\cite{zoetekouw_prl06}.

\begin{figure}
\begin{center}
\includegraphics[width=0.6\columnwidth]{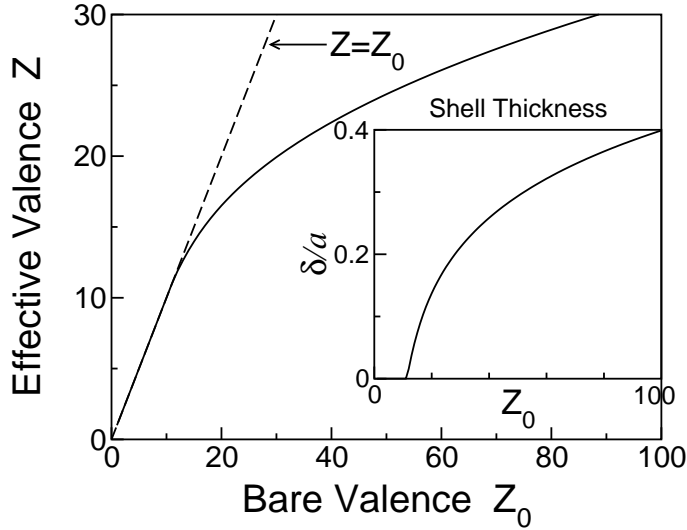}
\end{center}
\caption{\label{fig-crz}
Effective valence $Z$ vs.~bare valence $Z_0$ for a deionized suspension
($c_s\simeq 0$) of macroions of radius $a=2$ nm and volume fraction $\eta=0.01$.
Inset: counterion association shell emerges and thickens beyond threshold $Z_0$.
}
\end{figure}

Figure~\ref{fig-crz} illustrates the distinction between the bare and effective
macroion valences.  For a sufficiently small bare valence, Eq.~(\ref{delta2})
has no real-valued solution for any nonzero association shell thickness.
In this case, all counterions are free ($\delta=0$, $v_s=0$) and the free energy
is minimized by $Z=Z_0$.  Beyond a threshold bare valence, however, the association
shell appears and rapidly thickens with increasing $Z_0$.  Correspondingly, the
free energy minimum shifts to $Z<Z_0$.  With increasing $Z_0$, the effective
valence grows logarithmically, in contrast to the saturation observed for
polyelectrolytes~\cite{manning69,Oosawa} and predicted for colloidal suspensions
by PB cell theories~\cite{alexander84,belloni98,trizac02} and Debye-H\"uckel theories~\cite{levin98,tamashiro98,levin01,levin03,trizac04}.

\begin{figure}
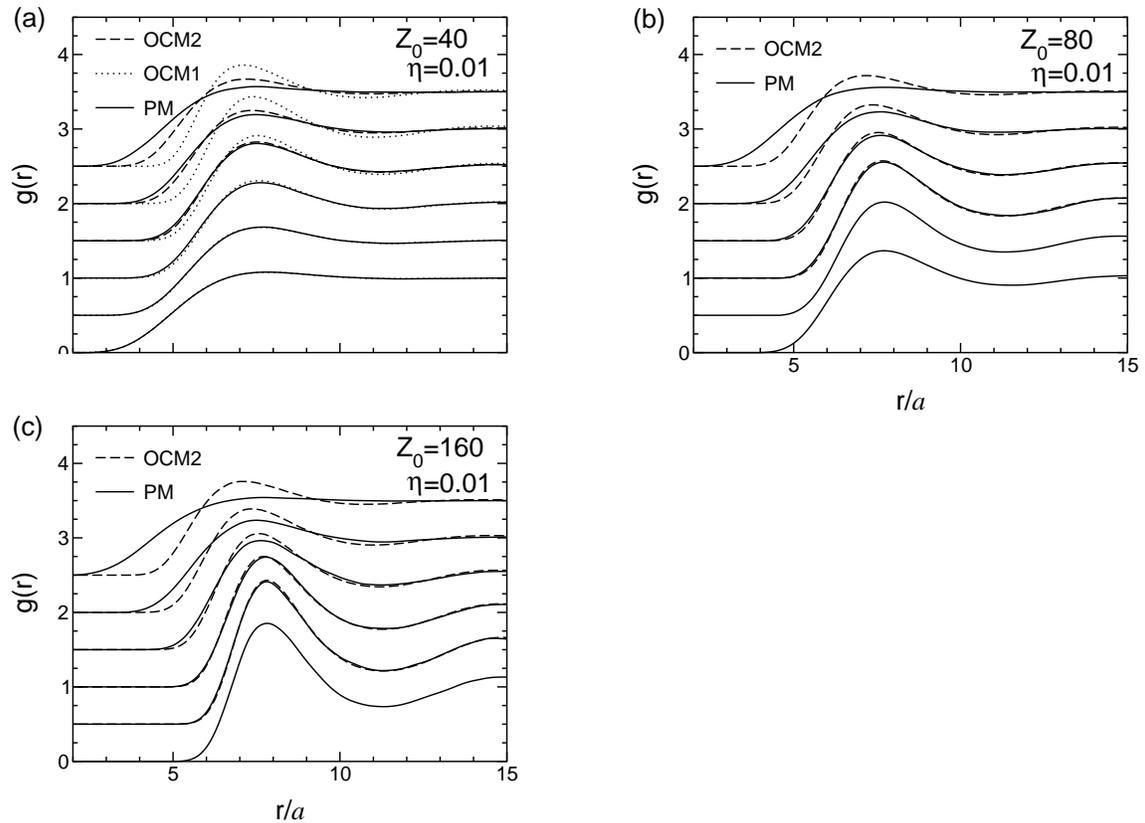

\includegraphics[width=0.45\columnwidth]{gr.z40.e01.eps}
\includegraphics[width=0.45\columnwidth]{gr.z80.e01.eps}
\includegraphics[width=0.45\columnwidth]{gr.z160.e01.eps}
\caption{\label{fig-gr1}
Macroion-macroion pair distribution function $g(r)$ vs. radial distance $r$
(units of macroion radius $a$) of salt-free suspensions computed from
Monte Carlo simulations of the effective one-component model, with
(OCM2, dashed curves) and without (OCM1, dotted curves) charge renormalization,
and of the primitive model (PM)~\cite{linse00} with explicit counterions
(solid curves) for volume fraction $\eta=0.01$, electrostatic coupling
parameters $\Gamma=\lambda_B/a=$ 0.0222, 0.0445, 0.0889, 0.1779, 0.3558, 0.7115
(bottom to top), and bare macroion valence $Z_0=40$ (a), 80 (b), and 160 (c).
For clarity, curves are vertically offset in steps of 0.5.
}
\end{figure}

\begin{figure}
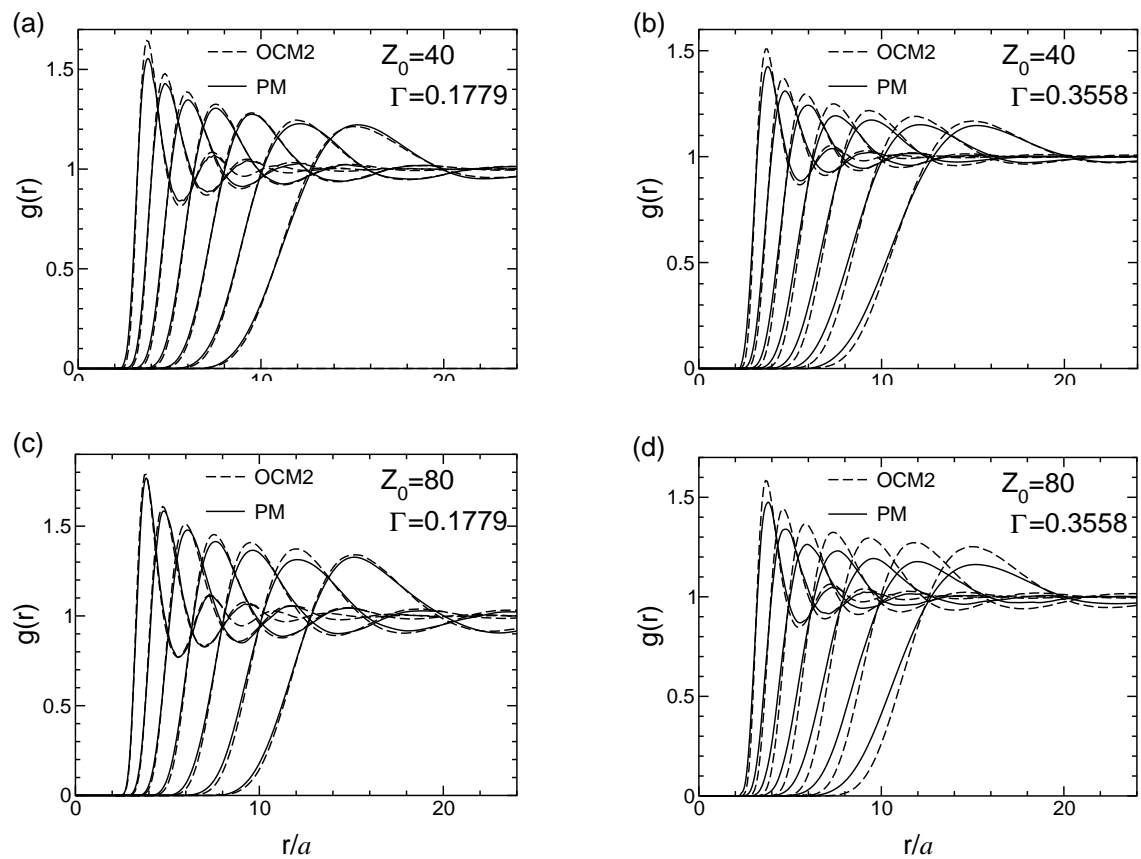

\includegraphics[width=0.45\columnwidth]{gr.z40.g1779.eps}
\includegraphics[width=0.45\columnwidth]{gr.z40.g3558.eps}
\includegraphics[width=0.45\columnwidth]{gr.z80.g1779.eps}
\hspace*{1.3cm}
\includegraphics[width=0.45\columnwidth]{gr.z80.g3558.eps}
\caption{\label{fig-gr2}
Macroion-macroion pair distribution function $g(r)$ vs. radial distance $r$
(units of macroion radius $a$) of salt-free suspensions computed from
Monte Carlo simulations of the charge-renormalized effective one-component model
(OCM2, dashed curves) and of the primitive model~\cite{linse00}
(PM, solid curves) for various bare macroion valences $Z_0$,
electrostatic coupling parameters $\Gamma=\lambda_B/a$, and volume fractions
$\eta=$ 0.00125, 0.0025, 0.005, 0.01, 0.02, 0.04, and 0.08 (right to left).
}
\end{figure}

We first test the capacity of the charge renormalization theory~\cite{denton08}
to predict the macroion structure.  Figures~\ref{fig-gr1} and \ref{fig-gr2}
compare results for the pair distribution function $g(r)$ from our simulations
of the OCM, using charge-renormalized effective interactions as input, and
corresponding data from extensive Monte Carlo simulations of the primitive model
(PM)~\cite{linse00} for salt-free suspensions over ranges of bare valence,
volume fraction, and electrostatic coupling strength $\Gamma=\lambda_B/a$.
Well beyond the threshold for charge renormalization ($Z_0\Gamma\simeq 7$),
the OCM and PM results agree closely.  Significant deviations emerge only for
$Z_0\Gamma>15$, the OCM consistently overpredicting the macroion structure.
For reference, Fig.~\ref{fig-gr1} also includes OCM results obtained with
unrenormalized effective interactions, illustrating the impact of
charge renormalization on strongly coupled suspensions.

\begin{figure}
\begin{center}
\includegraphics[width=0.7\columnwidth]{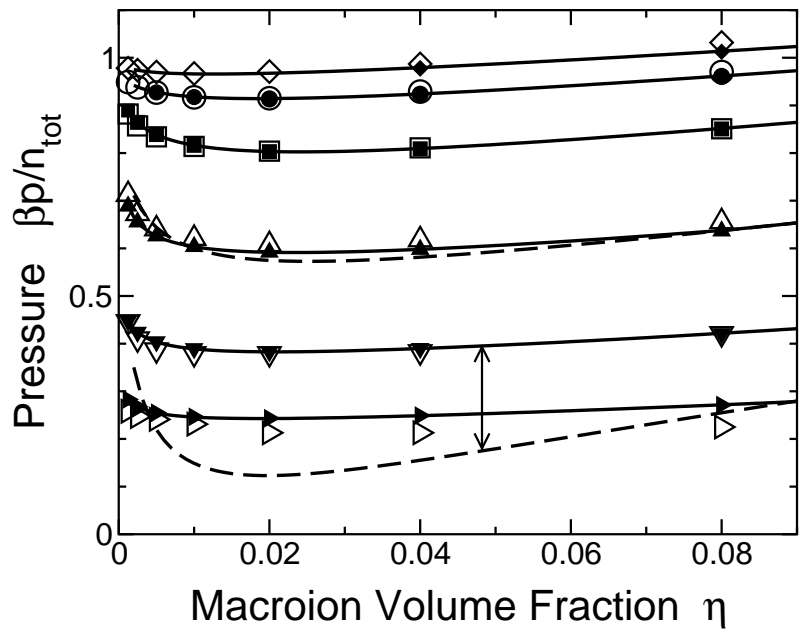}
\end{center}
\caption{\label{fig-crpe}
Total reduced pressure $\beta p/n_{\rm tot}$ vs.~macroion volume fraction $\eta$,
where $n_{\rm tot}=(Z_0+1)n_m$ (total ion density), of salt-free suspensions
with bare macroion valence $Z_0=40$ and electrostatic coupling constants
(from top to bottom) $\Gamma=$ 0.0222, 0.0445, 0.0889, 0.1779, 0.3558, 0.7115.
Open symbols: Monte Carlo simulations of the primitive model~\cite{linse00}.
Filled symbols: Monte Carlo simulations of the effective one-component model.
(Symbol sizes exceed error bars.)  Curves: variational theory with
(solid) and without (dashed) charge renormalization.
The double-ended arrow points to corresponding curves for $\Gamma=0.3558$.
The dashed curve for $\Gamma=0.7115$ is off-scale, the pressure being negative.
}
\end{figure}

\begin{figure}
\begin{center}
\includegraphics[width=0.7\columnwidth]{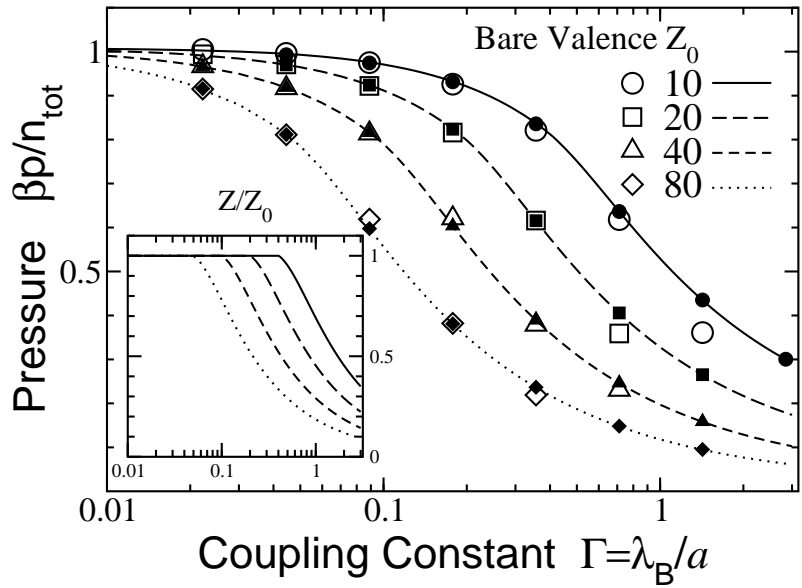}
\end{center}
\caption{\label{fig-crpg}
Total reduced pressure $\beta p/n_{\rm tot}$ vs.~electrostatic coupling parameter
$\Gamma$ of salt-free suspensions with fixed volume fraction $\eta=0.01$ and
bare macroion valence (top to bottom) $Z_0=$ 10, 20, 40, 80.
Open symbols: Monte Carlo simulations of the primitive model~\cite{linse00}.
Filled symbols: Monte Carlo simulations of the effective one-component model.
(Symbol sizes exceed error bars.)  Curves: charge-renormalized variational theory.
Inset: Ratio of effective to bare macroion valence $Z/Z_0$ vs. $\Gamma$.
}
\end{figure}

Finally, we test predictions of the charge renormalization theory~\cite{denton08}
for thermodynamics.  Figures~\ref{fig-crpe} and \ref{fig-crpg} directly compare the
pressures computed from our OCM simulations [via Eqs.~(\ref{pvirial})-(\ref{ptail})]
with corresponding data from primitive model simulations~\cite{linse00} of
salt-free suspensions over ranges of bare valence, volume fraction, and
electrostatic coupling strength.  The OCM and primitive model results are seen to be
in excellent agreement up to the highest couplings for which primitive model data are
available ($Z_0\Gamma\simeq 30$), well beyond the charge renormalization threshold.
Figures~\ref{fig-crpe} and \ref{fig-crpg} also compare the pressures resulting from
our OCM simulations with predictions of the variational theory [Eq.~(\ref{ptot})].
The near exact agreement between simulation and theory for the OCM confirms the
accuracy of the variational approximation for the macroion free energy [Eq.~(\ref{fm})]
and demonstrates the equivalence of the thermodynamic and virial expressions for the
pressure, i.e., Eqs.~(\ref{ptot}) and (\ref{pvirial}), respectively.

\section{Conclusions}\label{conclusions}

Summarizing, we have performed Monte Carlo simulations of the one-component model
of charged colloids, using as input effective electrostatic interactions predicted
by a recently proposed charge renormalization theory~\cite{denton08}.  Structural
and thermodynamic properties of salt-free suspensions are computed and directly
compared with simulations of the primitive model.  For bare valences $Z_0$ and
electrostatic coupling strengths $\Gamma$ considerably exceeding the renormalization
threshold, the pair distribution function and pressure are in close agreement with
corresponding results from simulations of the primitive model.
Significant discrepancies between the OCM and PM simulations are observed in $g(r)$
for $Z_0\Gamma>15$, and in the pressure only for $Z_0\Gamma>30$.
The level of agreement appears to be comparable to the charge-renormalization scheme
of ref.~\cite{lobaskin-linse99}, which is based on a PB cell model~\cite{alexander84}.
The computationally practical approach described here may provide a useful modeling
alternative, for some applications, to molecular-scale simulations of charged
colloids.  Further comparisons with primitive model simulation data, particularly
at nonzero salt concentrations, would help to chart the validity range of
the charge-renormalized OCM.

The threshold for charge renormalization closely coincides with the onset of
a spinodal phase instability predicted by linearized effective-interaction
theories~\cite{denton06,vRH97,vRDH99,warren} for deionized (but not salt-free)
suspensions of highly charged colloids.  Although observations consistent with bulk
phase separation have been reported~\cite{tata92,ise94,ise96,tata97,ise99,matsuoka,
grier97,groehn00}, the experimental picture of deionized suspensions remains cloudy.
Furthermore, the relevant parameter regime is not yet accessible to primitive model
simulations.  Recent studies based on the PB cell model have concluded that
the predicted instability may be an artifact of linearization
approximations~\cite{klein01,deserno02,tamashiro03}.  Preliminary calculations based
on the present approach indicate, however, that the phase instability may survive
charge renormalization~\cite{denton-cr08}.  Future work will continue to explore
the remarkable phase behavior of deionized suspensions.

\section*{Acknowledgments}
This work was supported by the National Science Foundation (DMR-0204020)
and the Petroleum Research Fund (PRF 44365-AC7).  We thank Per Linse for
helpful correspondence and for providing primitive model simulation data.

\end{document}